\documentclass[preprints,article,submit,oneauthor]{Definitions/mdpi}
\makeatletter
\AtBeginDocument{%
  \def\today{September 2, 2026}%
  \rfoot{}%
  \lfoot{\fontsize{8}{8}\selectfont Preprint version of \today}%
  \fancypagestyle{plain}{%
    \fancyhf{}%
    \fancyhead{\null\vspace{8pt}}%
    \fancyfoot[L]{\fontsize{8}{8}\selectfont Preprint version of \today}%
    \fancyfoot[R]{\fontsize{8}{8}\selectfont 1 of \pageref*{LastPage}}%
  }%
}
\makeatother
\makeatletter
\AtBeginDocument{\let\linenumbers\relax}
\AfterEndPreamble{\nolinenumbers}
\makeatother
\usepackage[T1]{fontenc}
\usepackage[utf8]{inputenc}
\usepackage{amssymb}
\usepackage{longtable}
\usepackage{array}
\usepackage{calc}
\usepackage{etoolbox}
\makeatletter
\patchcmd\longtable{\par}{\if@noskipsec\mbox{}\fi\par}{}{}
\makeatother

\AtBeginEnvironment{longtable}{\footnotesize\setlength{\tabcolsep}{6pt}}

\firstpage{1}
\pubvolume{1}
\issuenum{1}
\articlenumber{0}
\pubyear{2026}
\copyrightyear{2026}
\datereceived{ }
\daterevised{ }
\dateaccepted{ }
\datepublished{ }

\Title{Statevector-to-Hardware Reconstruction of a Four-Qubit ZZ Quantum Kernel: A Single-Backend Case Study of Three Execution Jobs}
\Author{Rostyslav Sipakov}
\AuthorNames{Rostyslav Sipakov}
\address{Department of Environmental Protection and Occupational Safety Technologies, Kyiv National University of Construction and Architecture, Povitrianykh Syl, 31, Kyiv 03037, Ukraine}
\corres{Correspondence: sipakov.rv@knuba.edu.ua}

\abstract{Hardware noise and finite sampling perturb the fidelity estimates
forming a quantum-kernel Gram matrix. We measured how far three
hardware-reconstructed Gram matrices depart from an exact statevector
reference for one frozen four-qubit ZZ feature map on \(N=24\) indoor
air-quality windows, executed on \texttt{ibm\_fez} at 1024 shots per
circuit in three single, non-interleaved jobs: baseline, dynamical
decoupling alone, and gate twirling alone. All were complete, finite,
and positive-semidefinite. Off-diagonal root-mean-squared error (RMSE)
against the reference was 0.0878, 0.0864, and 0.0427; full-matrix
centered kernel alignment (CKA) ranged 0.933--0.989 and the post hoc
diagonal-excluded (U-centered) CKA 0.816--0.986. The gate-twirled job
deviated least on every reported geometry axis; its baseline contrasts
are deletion-stable for the Spearman, mean-absolute-error, RMSE, and
full-matrix CKA diagnostics, while the Pearson and diagonal-excluded
contrasts fall just below that convention. Dynamical decoupling was not
separated from the baseline. The observed error exceeded both
finite-shot reference scales, so, under those sampling-only models,
sampling does not explain it. Centered kernel--target alignment did not
track reconstruction fidelity and stayed at or below each
label-permutation reference: implementation fidelity and task relevance
are distinct diagnostic axes. All configuration-level statements
describe three realized jobs on one backend; no mitigation-efficacy,
classifier-superiority, forecasting, or quantum-advantage claim is made.}


\begin{document}
\hypertarget{introduction}{%
\section{Introduction}\label{introduction}}

\hypertarget{background-and-motivation}{%
\subsection{Background and motivation}\label{background-and-motivation}}

A quantum feature map embeds a classical input \(x\) into a quantum
state, \(x \mapsto \lvert\phi(x)\rangle\), and the associated quantum
kernel measures similarity through a squared overlap,

\[
K(x_i,x_j)
{}=
\bigl\lvert\langle\phi(x_i)\mid\phi(x_j)\rangle\bigr\rvert^{2},
\tag{1}
\]

so that the kernel (Gram) matrix encodes the feature-map-induced
pairwise-similarity structure of the dataset
\citep{havlicek2019supervised, schuld2019quantum, schuld2021supervised}.
Every downstream kernel method sees the data only through this matrix,
so the fidelity of the matrix itself is a precondition for any learning
claim built on it. On near-term hardware, finite sampling and device
noise perturb the estimated overlaps, driving kernel entries toward a
common background and flattening the matrix relative to its intended
form
\citep{thanasilp2024exponential, heyraud2022noisy, wang2021towards}.

Our data come from indoor air-quality (IAQ) monitoring with low-cost
duplicate sensors, whose noisy, partially redundant multivariate streams
make a difficult real-world test case for similarity-based learning
\citep{morawska2018applications, karagulian2019review}. The broader
monitoring project concerns redundancy- and missingness-aware (RMA)
sensor modeling; the present hardware experiment is deliberately
restricted to a single fixed four-qubit ZZ feature map (ZZ4) and does
not execute any RMA kernel on hardware. This paper does not ask whether
a quantum computer improves indoor air-quality prediction. It asks the
logically prior question: can the statevector ZZ4 kernel be
reconstructed on real IBM Quantum hardware with interpretable, bounded
distortion? Characterizing how a designed kernel is changed by execution
is a prerequisite for, and distinct from, any later claim of quantum
advantage
\citep{peters2021machine, hubregtsen2022training, glick2024covariant, huang2021power}.

\hypertarget{related-work}{%
\subsection{Related work}\label{related-work}}

The ZZ-type feature map and the squared-overlap quantum kernel originate
in supervised learning with quantum-enhanced feature spaces
\citep{havlicek2019supervised}, with supervised quantum models later
identified as kernel methods
\citep{schuld2019quantum, schuld2021supervised}. Benchmarking shows that
quantum-kernel behavior hinges on encoding, bandwidth, and concentration
rather than on the mere use of a quantum device
\citep{schnabel2025quantum, gilfuster2024expressivity}: without the
right inductive bias, generalization can degrade
\citep{kubler2021inductive}, kernel values can concentrate toward a
constant \citep{thanasilp2024exponential}, and sufficient classical data
can erase the advantage \citep{huang2021power}. Concentration is
therefore not only a hardware-noise phenomenon: data-scaling
(bandwidth-type) hyperparameters control the noiseless spectrum and
inductive bias of the kernel itself \citep{canatar2023bandwidth}, and
the usefulness of a quantum kernel is a separate question from its
faithful execution \citep{incudini2025toward}. These results motivate
treating the statevector kernel, not the hardware kernel, as the
intended object, and asking how faithfully it is reconstructed in
execution.

On noisy hardware, quantum-kernel classifiers have been demonstrated at
the 17--27-qubit scale \citep{peters2021machine, glick2024covariant} and
embedding kernels have been trained directly on devices
\citep{hubregtsen2022training}, but finite-shot estimation, noise, and
compilation all degrade the estimated kernel
\citep{wang2021towards, thanasilp2024exponential}, decoherence reduces
its effective rank \citep{heyraud2022noisy}, and the realized benefit of
mitigation depends jointly on device and compilation
\citep{ji2024synergistic, cai2023quantum}. The execution configurations
compared here, dynamical decoupling
\citep{viola1999dynamical, ezzell2023dynamical} and gate (Pauli)
twirling \citep{wallman2016noise}, are standard runtime-selectable
mitigation strategies within this toolbox
\citep{temme2017error, hicks2022active, smith2021qubit}. The closest
hardware-fidelity neighbor validates a quantum-kernel support vector
machine on the same backend family and reports high
statevector-to-hardware fidelity correlation with overly flat quantum
eigenspectra \citep{kakavand2026benchmarking} (preprint); studies that
combine mitigations and judge them by downstream accuracy
\citep{singh2026benchmarking} answer a different question from
kernel-geometry reconstruction \citep{cerezo2022challenges}.

We measure reconstruction fidelity with representation-similarity and
alignment tools: centered kernel alignment (CKA) between two kernels
\citep{kornblith2019similarity}, centered (Cortes-type) kernel--target
alignment (KTA) between a kernel and the label Gram matrix
\citep{cortes2012algorithms, hubregtsen2022training}, entropy-based
effective rank \citep{roy2007effective, heyraud2022noisy}, and
positive-semidefinite (PSD) diagnostics \citep{higham1988computing}.
Both CKA and centered KTA are invariant to affine (scale-and-shift)
changes of a kernel, so a purely depolarizing-type contraction cannot
move them; a recent kernel-level comparison instead reads depolarizing
noise as increasing quantum--classical CKA \citep{rza2026beyond}
(preprint). This sets up the central tension of the study: alignment
with the statevector geometry and alignment with the labels need not
move in the same direction.

\hypertarget{research-gap-and-contribution}{%
\subsection{Research gap and
contribution}\label{research-gap-and-contribution}}

Quantum kernels have been executed on hardware, and the individual
mitigation strategies and geometry diagnostics are each well
established. What is comparatively underexplored is their combination on
a single, fixed kernel: a fixed-protocol, statevector-referenced
measurement of reconstruction fidelity for one frozen kernel on a real
device, compared across execution configurations and read against an
explicit finite-shot reference scale
\citep{shastry2022shot, wang2021towards}. Neither CKA nor KTA is
introduced here. The contribution is their paired,
statevector-referenced use as separate implementation-fidelity and
task-relevance axes within an artifact-grounded hardware-kernel
diagnostic case study: entrywise error and CKA answer how faithfully the
designed Gram matrix was reconstructed; centered KTA against a
permutation reference answers whether the kernel's structure is aligned
with the task; a finite-shot reference scale bounds what sampling alone
can explain; and none of these answers substitutes for the others, or
for downstream accuracy. Concretely, we (i) execute the fixed ZZ4 kernel
on IBM Quantum hardware for a frozen subset of \(N=24\) IAQ windows;
(ii) compare baseline, dynamical-decoupling, and gate-twirling
configurations, each submitted as a single non-interleaved job; (iii)
report rank-order, linear, entrywise, alignment (full-matrix and
diagonal-excluded CKA; centered KTA), spectral, and PSD diagnostics
against the statevector reference; (iv) document a CKA/KTA tension in
which the most faithful reconstruction does not carry the highest label
alignment; and (v) compare the observed off-diagonal root-mean-squared
error (RMSE) with conservative global and entry-resolved finite-shot
reference scales. These contributions are diagnostic: the single-job
design supports descriptive comparison, not causal mitigation-efficacy
estimates, and the fixed \(N=24\) subset supports
reconstruction-fidelity claims, not predictive ones.

\hypertarget{research-questions-and-scope}{%
\subsection{Research questions and
scope}\label{research-questions-and-scope}}

This work is a fixed-subset, single-backend hardware-kernel case study
of three execution jobs (\texttt{ibm\_fez}, 1024 shots per circuit),
designated \emph{Wave 1} in the project's decision record. It does not
estimate general mitigation efficacy and does not test IAQ forecasting
accuracy, quantum advantage, or hardware classifier superiority.
``Survival'' is reported as a continuous, multi-metric description; no
binary pass/fail threshold was pre-specified. Directional expectations
recorded alongside the research questions were not independently
timestamped before the hardware results were available and are not
treated as confirmatory hypotheses (Supplementary Note S1). We address
four research questions.

\textbf{RQ1 (geometry survival).} To what extent is the geometry of the
fixed four-qubit ZZ4 statevector kernel preserved when reconstructed on
IBM Quantum hardware for the frozen \(N=24\) subset, as measured by
rank-order (Spearman), linear (Pearson), entrywise (mean,
root-mean-squared, and maximum absolute error), and centered full-matrix
(CKA) agreement with the statevector reference?

\textbf{RQ2 (configuration differences).} What descriptive differences
in centered-geometry preservation and off-diagonal entrywise distortion
are observed among the three configurations (baseline, dynamical
decoupling alone, and gate twirling alone), each submitted as a single
non-interleaved job?

\textbf{RQ3 (finite-shot reference scale).} How does the observed
off-diagonal hardware-versus-statevector RMSE compare with a
conservative global and an entry-resolved finite-shot reference scale
under the Wave 1 reconstruction?

\textbf{RQ4 (label alignment as a distortion diagnostic).} Does the
intended ZZ4 statevector kernel align with the frozen event-onset labels
beyond a random-label permutation reference on this subset, and should
any CKA/KTA divergence be interpreted as a distortion diagnostic rather
than as evidence of supervised predictive improvement?

Throughout, statevector and hardware quantities are kept distinct, and
all alignment and spectral changes are treated as diagnostics of
hardware distortion. Section 2 specifies the data, kernel, hardware
protocol, and metrics; Section 3 reports the results; Section 4
discusses implications; Section 5 consolidates limitations; Section 6
concludes. Full protocol, derivation, and extended diagnostic detail is
relocated to the Supplementary Methods (location map in Table SM.0) and
Supplementary Notes S1--S3.

\hypertarget{materials-and-methods}{%
\section{Materials and methods}\label{materials-and-methods}}

\hypertarget{data-and-frozen-subset}{%
\subsection{Data and frozen subset}\label{data-and-frozen-subset}}

The data are real indoor air-quality duplicate-sensor records organized
as a forecasting dataset (30-minute window stride, one-hour horizon).
The binary target \texttt{y\_event\_onset\_next\_1h} indicates whether a
new air-quality event starts within the next hour; eligible windows
exclude already-active events and invalid future labels. The feature set
\texttt{F\_quantum\_4} comprises four one-hour pollutant summaries
(\texttt{pm25\_mean\_last\_1h}, \texttt{pm10\_mean\_last\_1h},
\texttt{hcho\_mean\_last\_1h}, \texttt{tvoc\_mean\_last\_1h}), imputed
from training data, min-max scaled to \([0,\pi]\) under a train-only
policy, and clipped out of range.

A subset of \(N=24\) observation windows (16 training, 8 test; 12
event-onset and 12 non-onset) was selected and frozen before hardware
execution was authorized, with dated, checksum-locked freeze artifacts
preceding job creation. The windows span 2024-12-22 to 2026-04-20 with
median spacing \(\approx 4.7\) days (per-window ledger in Supplementary
Table S2.9). The frozen subset yields \(N(N+1)/2=300\) unordered kernel
evaluations (276 unique off-diagonal pairs), enumerated in a fixed,
label-blind pair inventory (Supplementary Methods A1). The governing
decision record states \texttt{STOP\_AFTER\_WAVE1\_REPORT\_RESULTS}: the
subset is not an adjustable analysis input, and no window was added,
removed, or reweighted after authorization (Supplementary Methods
A1--A4). The public Wave 1 reproducibility package at a fixed commit
(github.com/rsipakov/iaq-quantum-kernel-wave1-reproducibility, release
\texttt{v1.2-wave1-manuscript}; Zenodo DOI 10.5281/zenodo.21332398)
contains the frozen subset, the statevector reference, the raw and
reconstructed hardware artifacts, and the analysis scripts;
Supplementary Note S1 maps every reported quantity to its artifact.

\hypertarget{zz4-feature-map-statevector-reference-and-hardware-estimator}{%
\subsection{ZZ4 feature map, statevector reference, and hardware
estimator}\label{zz4-feature-map-statevector-reference-and-hardware-estimator}}

ZZ4 is Qiskit's \texttt{ZZFeatureMap} on four qubits (one per feature):
two repetitions of a Hadamard layer followed by a data-encoding layer of
single- and two-qubit \(Z\) rotations, with linear (nearest-neighbor)
entanglement and the default scaling \(\alpha=2\)
\citep{havlicek2019supervised}. The configuration was fixed before
execution and never re-tuned. The statevector reference is the exact
squared-fidelity kernel of this map,

\[
K_{\mathrm{SV}}(i,j)
{}=
\bigl\lvert\langle\phi(\tilde{x}_i)\mid\phi(\tilde{x}_j)\rangle\bigr\rvert^{2},
\tag{2}
\]

computed for all 24 windows; its diagonal equals unity by construction.
Each hardware entry was estimated with a compute--uncompute fidelity
circuit,
\(U_{\mathrm{ZZ4}}(\tilde{x}_j)^{\dagger}U_{\mathrm{ZZ4}}(\tilde{x}_i)\)
applied to \(\lvert 0\rangle^{\otimes 4}\), whose all-zero outcome
probability equals \(K_{\mathrm{SV}}(i,j)\) in the noiseless limit. On
hardware the entry is the finite-shot estimator

\[
\widehat{K}_r(i,j)
{}=
\frac{n_{0^4,r}(i,j)}{N_r(i,j)},
\tag{3}
\]

with \(n_{0^4,r}\) the observed all-zero count and \(N_r=1024\) the
observed shots, for regime
\(r\in\{\mathrm{H0},\mathrm{H1},\mathrm{H2}\}\). Symmetrization mirrors
the measured upper triangle; the measured diagonal is retained rather
than forced to one (mean \(\approx 0.94\); diagonal circuits reduce to
measurement-only bodies, so these entries mainly reflect preparation,
readout, and finite-shot effects). All three reconstructed matrices are
complete (\(576\) finite entries) and already positive semidefinite
(uncorrected \(\lambda_{\min}=0.429\), \(0.462\), \(0.232\); statevector
\(0.0565\)), so no reported metric depends on PSD replacement (audit in
Supplementary Methods A4).

\hypertarget{hardware-protocol-and-execution-configurations}{%
\subsection{Hardware protocol and execution
configurations}\label{hardware-protocol-and-execution-configurations}}

The pilot ran on the 156-qubit IBM Quantum backend \texttt{ibm\_fez}
with the \texttt{SamplerV2} primitive \citep{javadi-abhari2024qiskit}
(Qiskit 2.4.1, qiskit-ibm-runtime 0.46.1) in backend job mode, one job
per configuration, each covering all 300 pair circuits:
\texttt{d7vf6n3ack5s73bfc0eg} (\texttt{H0}, baseline),
\texttt{d7vf8ocinasc738u1bhg} (\texttt{H1}, dynamical decoupling only,
\texttt{XX} sequence), and \texttt{d7vfbsfmrars73d84u20} (\texttt{H2},
gate twirling only, \texttt{active-accum}, resolved to 16 randomizations
per circuit). All three jobs completed (\texttt{DONE}) with 300
retrieved PUB results each and no retrieval failure, and they ran
sequentially within one \(\approx 13.3\)-minute window on 2026-05-09
(UTC); the persisted job manifest records their submission at 08:42:05,
08:46:26, and 08:53:05 UTC (\texttt{H0}, \texttt{H1}, \texttt{H2}, in
that order), for \(921{,}600\) observed shots and 244 billed QPU seconds
in total. The 300 logical circuit bodies are shared across regimes; the
regimes differ only in Sampler-level runtime options (SHA-256-locked
records). Compiled depth was at most 102 with at most 22 two-qubit gates
(\texttt{cz}), at \texttt{optimization\_level=1}; the physical-qubit
layout, transpilation seed, and a contemporaneous calibration table were
not persisted. The manuscript refers to the artifact regimes
\texttt{H0}/\texttt{H1}/\texttt{H2} as configurations
\texttt{M0}/\texttt{M1}/\texttt{M2} where convenient; the alias is
notational only (Supplementary Methods A3). Shots were a budget-safe
1024 per circuit (planned: 4096). For \texttt{H2}, the 1024 shots are
pooled across the 16 twirling randomizations; per-randomization counts
were not persisted, so between-randomization variability cannot be
estimated from the recorded data (Supplementary Methods A2). No combined
decoupling-plus-twirling configuration was executed.

\hypertarget{geometry-and-task-alignment-metrics}{%
\subsection{Geometry and task-alignment
metrics}\label{geometry-and-task-alignment-metrics}}

Scalar agreement and entrywise-error metrics are evaluated on the
off-diagonal domain \(\Omega=\{(i,j): i\neq j\}\)
(\(\lvert\Omega\rvert=552\) directed entries; equivalent to the 276
unordered pairs by symmetry): Spearman and Pearson correlations with the
statevector entries, and mean, root-mean-squared, median, and maximum
absolute error. Matrix-level metrics retain the full symmetric matrix
including the measured diagonal: the centered kernel alignment

\[
\operatorname{CKA}(K_m,K_{\mathrm{SV}})
{}=
\frac{\langle HK_mH,\,HK_{\mathrm{SV}}H\rangle_F}
{\lVert HK_mH\rVert_F\,\lVert HK_{\mathrm{SV}}H\rVert_F},
\qquad
H=I_N-\tfrac{1}{N}\mathbf{1}\mathbf{1}^{\top},
\tag{4}
\]

the centered kernel--target alignment
\(\operatorname{KTA}_{\mathrm{c}}(K_m,y)=\operatorname{CKA}(K_m,yy^{\top})\)
with balanced signed labels \(y_i\in\{-1,+1\}\), and the entropy
effective rank of the zero-clipped spectrum. Both centered functionals
are invariant to the affine map
\(K\mapsto aK+b\,\mathbf{1}\mathbf{1}^{\top}\) (\(a>0\)), so a purely
depolarizing contraction cannot move them; any positive hardware KTA
uplift certifies a non-affine distortion component, without by itself
identifying a mechanism (Supplementary Methods A5--A8). Because the
shared near-unit diagonal inflates the full-matrix CKA, a post hoc
diagonal-excluded (U-centered) CKA computed from the HSIC\(_1\)
estimator \citep{song2012feature} is reported alongside it
(Supplementary Table S2.5). Statevector-referenced tension quantities
are the CKA loss
\(L_{\mathrm{CKA},r}=1-\operatorname{CKA}(K^{(r)}_{\mathrm{hw}},K_{\mathrm{SV}})\)
and the label uplift
\(\Delta_{\mathrm{KTA},r}=\operatorname{KTA}_{\mathrm{c}}(K^{(r)}_{\mathrm{hw}},y)-\operatorname{KTA}_{\mathrm{c}}(K_{\mathrm{SV}},y)\).

\hypertarget{statistical-policy-and-reference-scales}{%
\subsection{Statistical policy and reference
scales}\label{statistical-policy-and-reference-scales}}

The statistical unit is the frozen observation window: kernel entries
sharing a window are dependent, so the 552 directed entries are
deterministic summary domains, not sampling units, and no entrywise
correlation-test \(p\)-values or bootstrap intervals are used.
Uncertainty statements are limited to leave-one-window-out jackknife
standard errors \citep{efron1981jackknife} and paired contrasts
preserving within-window covariance; the descriptive ratio
\(z_{\mathrm{desc}}=\Delta/\widehat{\mathrm{SE}}_{\mathrm{JK}}(\Delta)\)
is a scale-free stability diagnostic, not a test statistic, and is not
converted into \(p\)-values. For brevity, a contrast is called
\emph{deletion-stable} when \(\lvert z_{\mathrm{desc}}\rvert\ge 2\);
this is a readability convention conditional on the three realized jobs,
not a significance threshold, and it measures sensitivity to deletion of
frozen windows, not hardware run-to-run uncertainty. Non-resolution is
not equivalence. The persisted jackknife covers Spearman, Pearson, MAE,
full-matrix CKA, and centered KTA; deletion probes for RMSE, U-centered
CKA, and \(\Delta_{\mathrm{KTA}}\) were added at revision under the same
reading (Supplementary Methods A10).

Two finite-shot reference scales are defined at the executed \(S=1024\)
shots: the deliberately conservative global scale
\(\sigma_{\mathrm{ref,global}}=1/\sqrt{2S}\approx 0.022097\), and the
entry-resolved matrix-aware scale \(\sigma_{\mathrm{shot,matrix},r}\)
(root mean square of the per-entry binomial plug-ins over \(\Omega\),
\(\approx 0.0083\)). Both are diagnostic reference scales, not
uncertainty models or physical noise-model decompositions, and for
\texttt{H2} they use the pooled 1024-shot probabilities (Supplementary
Methods A9). Label alignment is referenced to fixed-seed
label-permutation nulls (\(B=5000\); statevector, with per-regime nulls
added at revision), a split-preserving permutation variant, a
count-level finite-shot resampling null (\(20{,}000\) replicates), and
two classical comparator kernels; the revision additions are
deterministic or fixed-seed recomputations from the frozen \texttt{v1.2}
artifacts, are explicitly post hoc, and none is a confirmatory test
(Supplementary Methods A10). The directional expectations E1--E4
recorded alongside the research questions were not independently
timestamped before the hardware results were available, are not treated
as confirmatory, and are reproduced verbatim, with their provenance
status, in Supplementary Note S1 (Section S1.5). All statistical outputs
are fixed-subset, post-reconstruction diagnostics; they support
kernel-geometry and distortion statements, not classifier accuracy,
hardware superiority, or quantum advantage.

Generative AI tools were used during preparation (figure-script review,
readability editing, reference verification, and an independent audit of
the public package); they were not used to design the study, execute
jobs, choose methodology, or produce any reported quantity, and all
reported quantities regenerate from the persisted pipeline (Code
availability).

\hypertarget{results}{%
\section{Results}\label{results}}

\hypertarget{execution-and-reconstruction}{%
\subsection{Execution and
reconstruction}\label{execution-and-reconstruction}}

All three configurations completed as single backend-mode jobs on
\texttt{ibm\_fez} at 1024 shots per circuit, with 300 retrieved PUB
results per regime, no retrieval failure, and complete, finite,
positive-semidefinite \(24\times 24\) reconstructions under the
measured-diagonal policy (Section 2.2; job identifiers in Section 2.3;
full execution ledger, resource accounting, and PSD audit in
Supplementary Methods B1 and A4). Every frozen quantity below traces to
a versioned, checksum-registered artifact of the public package; the
revision-added diagnostics are fixed-seed recomputations from those
artifacts, registered in release \texttt{v1.3-revision-diagnostics}
(Supplementary Notes S1--S2).

\hypertarget{reconstruction-fidelity-and-configuration-contrasts-rq1-rq2}{%
\subsection{Reconstruction fidelity and configuration contrasts (RQ1,
RQ2)}\label{reconstruction-fidelity-and-configuration-contrasts-rq1-rq2}}

Table 1 reports the headline distortion metrics; Figure 1 shows the four
Gram matrices and their entrywise errors.

\begingroup
\footnotesize
\setlength{\tabcolsep}{6pt}
\renewcommand{\arraystretch}{1.15}

\begin{longtable}[]{@{}lrrrrrrr@{}}
\toprule
\begin{minipage}[b]{0.15\columnwidth}\raggedright
Configuration\strut
\end{minipage} & \begin{minipage}[b]{0.09\columnwidth}\raggedleft
Artifact regime\strut
\end{minipage} & \begin{minipage}[b]{0.06\columnwidth}\raggedleft
Spearman\strut
\end{minipage} & \begin{minipage}[b]{0.06\columnwidth}\raggedleft
MAE\strut
\end{minipage} & \begin{minipage}[b]{0.06\columnwidth}\raggedleft
RMSE\strut
\end{minipage} & \begin{minipage}[b]{0.06\columnwidth}\raggedleft
CKA (full)\strut
\end{minipage} & \begin{minipage}[b]{0.10\columnwidth}\raggedleft
U-CKA (post hoc)\strut
\end{minipage} & \begin{minipage}[b]{0.20\columnwidth}\raggedleft
\(\operatorname{KTA}_{\mathrm{c}}\)\strut
\end{minipage}\tabularnewline
\midrule
\endhead
\begin{minipage}[t]{0.15\columnwidth}\raggedright
\texttt{M0} baseline\strut
\end{minipage} & \begin{minipage}[t]{0.09\columnwidth}\raggedleft
\texttt{H0}\strut
\end{minipage} & \begin{minipage}[t]{0.06\columnwidth}\raggedleft
\(0.741\)\strut
\end{minipage} & \begin{minipage}[t]{0.06\columnwidth}\raggedleft
\(0.0490\)\strut
\end{minipage} & \begin{minipage}[t]{0.06\columnwidth}\raggedleft
\(0.0878\)\strut
\end{minipage} & \begin{minipage}[t]{0.06\columnwidth}\raggedleft
\(0.933\)\strut
\end{minipage} & \begin{minipage}[t]{0.10\columnwidth}\raggedleft
\(0.8156\)\strut
\end{minipage} & \begin{minipage}[t]{0.20\columnwidth}\raggedleft
\(0.1833\)\strut
\end{minipage}\tabularnewline
\begin{minipage}[t]{0.15\columnwidth}\raggedright
\texttt{M1} dynamical decoupling\strut
\end{minipage} & \begin{minipage}[t]{0.09\columnwidth}\raggedleft
\texttt{H1}\strut
\end{minipage} & \begin{minipage}[t]{0.06\columnwidth}\raggedleft
\(0.775\)\strut
\end{minipage} & \begin{minipage}[t]{0.06\columnwidth}\raggedleft
\(0.0473\)\strut
\end{minipage} & \begin{minipage}[t]{0.06\columnwidth}\raggedleft
\(0.0864\)\strut
\end{minipage} & \begin{minipage}[t]{0.06\columnwidth}\raggedleft
\(0.937\)\strut
\end{minipage} & \begin{minipage}[t]{0.10\columnwidth}\raggedleft
\(0.8373\)\strut
\end{minipage} & \begin{minipage}[t]{0.20\columnwidth}\raggedleft
\(0.1815\)\strut
\end{minipage}\tabularnewline
\begin{minipage}[t]{0.15\columnwidth}\raggedright
\texttt{M2} gate twirling\strut
\end{minipage} & \begin{minipage}[t]{0.09\columnwidth}\raggedleft
\texttt{H2}\strut
\end{minipage} & \begin{minipage}[t]{0.06\columnwidth}\raggedleft
\(0.944\)\strut
\end{minipage} & \begin{minipage}[t]{0.06\columnwidth}\raggedleft
\(0.0257\)\strut
\end{minipage} & \begin{minipage}[t]{0.06\columnwidth}\raggedleft
\(0.0427\)\strut
\end{minipage} & \begin{minipage}[t]{0.06\columnwidth}\raggedleft
\(0.989\)\strut
\end{minipage} & \begin{minipage}[t]{0.10\columnwidth}\raggedleft
\(0.9863\)\strut
\end{minipage} & \begin{minipage}[t]{0.20\columnwidth}\raggedleft
\(0.1710\)\strut
\end{minipage}\tabularnewline
\bottomrule
\end{longtable}

\emph{Table 1. Headline statevector-to-hardware ZZ4 distortion metrics
for the three observed jobs.} Statevector references:
\(\operatorname{KTA}_{\mathrm{c}}(K_{\mathrm{SV}},y)=0.1585\);
\(\operatorname{erank}(K_{\mathrm{SV}})=17.97\) (hardware \(21.18\),
\(21.22\), \(19.79\)). Pearson correlations are \(0.827\), \(0.843\),
\(0.986\); median/maximum absolute errors \(0.0262/0.569\),
\(0.0261/0.564\), \(0.0162/0.264\). The U-centered CKA is the post hoc
diagonal-excluded (HSIC\(_1\)) variant, reported because the shared
near-unit measured diagonal (mean \(\approx 0.94\)) inflates the
full-matrix value. Uncertainty status (deletion scale, Supplementary
Methods B2): the \texttt{M2} contrasts against \texttt{M0} and
\texttt{M1} are deletion-stable for Spearman, MAE, RMSE, and full-matrix
CKA (\(\lvert z_{\mathrm{desc}}\rvert\approx 2.8\)--\(5\)); the
U-centered \texttt{M2-M0} contrast is weaker
(\(z_{\mathrm{desc}}=1.95\); \texttt{M2-M1} \(2.48\)); Pearson
\texttt{M2-M0} is borderline (\(1.92\)); no \texttt{M1-M0} contrast is
deletion-stable (\(\lvert z_{\mathrm{desc}}\rvert\le 1.96\)); all
centered-KTA contrasts are unresolved
(\(\lvert z_{\mathrm{desc}}\rvert\le 0.87\)). All values are point
estimates on the frozen \(N=24\) subset; no formal significance test or
mitigation-efficacy estimate is implied. Full and unsplit tables:
Supplementary Tables S2.1--S2.2, Supplementary Methods B2.

\endgroup

\textbf{Reconstruction fidelity (RQ1).} All three reconstructions
preserve the intended ZZ4 structure to a substantial descriptive degree.
Even the unmitigated baseline retains positive rank-order and linear
agreement (\(\rho=0.741\), \(r=0.827\)) and high full-matrix centered
alignment (\(0.933\); diagonal-excluded \(0.816\)). Preservation is
incomplete everywhere. The hardware kernels compress the off-diagonal
spread to \(28.9\%\), \(28.3\%\), and \(52.3\%\) of the statevector
off-diagonal variance (\(0.01866\)), the direction expected when noise
concentrates overlaps toward a common value
\citep{thanasilp2024exponential}. The worst-case entrywise error reaches
\(0.569\) (\texttt{H0}) and \(0.564\) (\texttt{H1}), more than half the
kernel range, against \(0.264\) for \texttt{H2}.

\textbf{Configuration contrasts (RQ2).} On every agreement metric the
observed ordering is gate twirling first, then dynamical decoupling,
then baseline. Relative to baseline, the gate-twirled job shows
\(47.5\%\) lower off-diagonal MAE, \(51.3\%\) lower RMSE, and a
centered-alignment loss \(1-\operatorname{CKA}\) of \(0.0113\) versus
\(0.0666\); on the diagonal-excluded variant the point separation widens
(\(0.9863\) versus \(0.8156\)) while its deletion contrast weakens
(\(z_{\mathrm{desc}}=1.95\) versus \(2.83\) full-matrix). The
baseline-to-decoupling differences are small and not deletion-stable on
any persisted metric. Effective rank is inflated on hardware (\(21.18\),
\(21.22\), \(19.79\) versus \(17.97\)), least under twirling; this is an
uncentered point-estimate diagnostic of spectral flattening,
complementary to the centered metrics (spectra in Supplementary Figure
S3.2; detail in Supplementary Methods B5). Among the three observed
jobs, the gate-twirled job is therefore the most faithful
reconstruction, and this is a descriptive statement about these jobs on
this backend, not a causal mitigation-efficacy estimate.

\begin{figure}[!htbp]
\centering
\includegraphics[width=\linewidth]{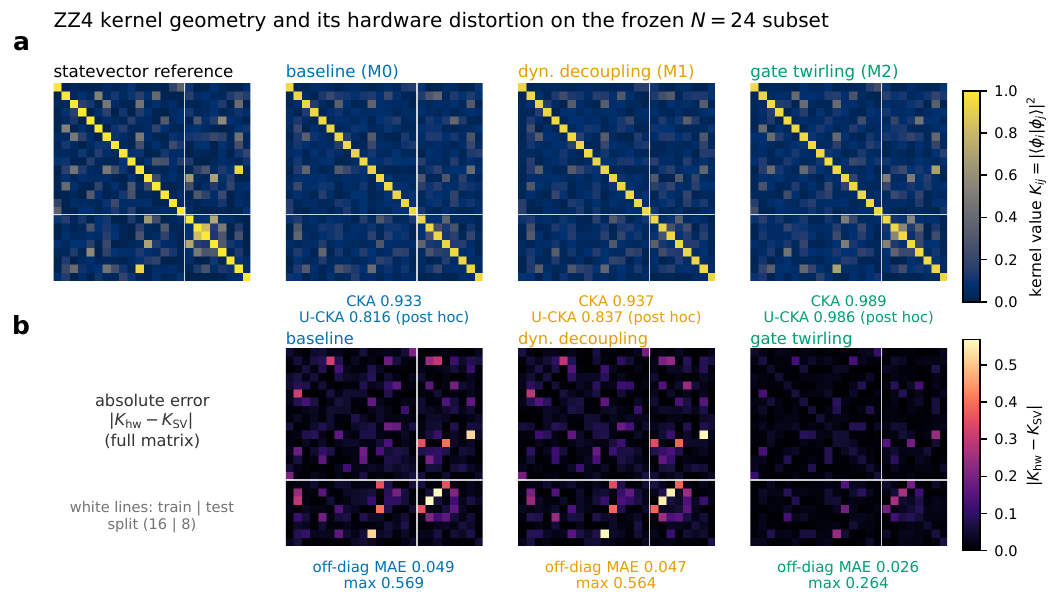}
\par\vspace{6pt}
\begin{minipage}{\linewidth}
\emph{Figure 1. ZZ4 kernel geometry and its hardware distortion on the
frozen \(N=24\) subset.} \emph{(a) The statevector Gram matrix and its
three hardware reconstructions on a shared \([0,1]\) scale; the bright
diagonal (mean \(\approx 0.94\)) is the measured hardware diagonal and
is not itself evidence of reconstruction fidelity. (b) Entrywise
absolute error \(\lvert K_{\mathrm{hw}}-K_{\mathrm{SV}}\rvert\); error
concentrates in a subset of window pairs under baseline and dynamical
decoupling and is markedly smaller in the gate-twirled job. White lines
mark the fixed \(16\,|\,8\) train/test split. Descriptive diagnostic:
single non-interleaved jobs on \texttt{ibm\_fez} at 1024 shots.}
\end{minipage}
\end{figure}

\hypertarget{task-alignment-and-the-ckakta-tension-rq4}{%
\subsection{Task alignment and the CKA/KTA tension
(RQ4)}\label{task-alignment-and-the-ckakta-tension-rq4}}

The statevector ZZ4 kernel does not align with the frozen event-onset
labels beyond a random-label reference: its centered alignment,
\(0.1585\), lies below the permutation-null mean (\(0.1710\)) with
\(p_{\mathrm{upper\text{-}tail}}=0.5988\) (\(B=5000\); seed-stable
across a 16-seed envelope, \([0.587,0.608]\)), and a split-preserving
permutation variant agrees (\(0.562\)). Two post hoc classical
comparators on the same inputs (linear and median-bandwidth RBF kernels)
likewise show no above-chance alignment (\(0.0076\) and \(0.0680\);
\(p=0.796\) and \(0.589\); Supplementary Table S2.8), consistent with
the weak alignment being a property of the frozen subset and target at
this scale rather than of ZZ4 specifically. This is an
absence-of-evidence statement at \(N=24\), not an equivalence claim, and
it rules out any IAQ forecasting claim from this pilot.

Against this reference, the hardware kernels show a small centered-KTA
\emph{uplift} (\(\Delta_{\mathrm{KTA}}=+0.0248\), \(+0.0230\),
\(+0.0125\) for \texttt{H0}, \texttt{H1}, \texttt{H2}), ordered opposite
to geometric fidelity: the most faithful job carries the lowest absolute
hardware KTA (\(0.1710\)), closest to the statevector value. Three
revision references bound its reading (Supplementary Table S2.6). First,
the uplift lies outside a count-level finite-shot resampling null around
the statevector probabilities (\(0/20{,}000\) replicates reach any
observed uplift; null SD \(0.0022\)), so it is not attributable to
finite-shot sampling under that reference. Second, every hardware kernel
sits at or below its own label-permutation mean
(\(p_{\mathrm{upper\text{-}tail}}=0.670\), \(0.711\), \(0.639\)), so the
uplift cannot be read as captured label signal. Third, a
numerator--denominator attribution locates it arithmetically: the
label-directed numerator \(y^{\top}Ky\) \emph{decreases} on hardware in
every configuration (\(-3.5\%\) to \(-5.6\%\), carried mainly by
measured-diagonal deflation), while the centered kernel norm decreases
substantially more (\(-16.6\%\), \(-16.8\%\), \(-12.5\%\)), so the
normalized alignment rises. Because the centered functionals annihilate
affine (depolarizing-type) contractions (Section 2.4), the uplift
certifies a non-affine distortion component whose action is
normalization-associated; it identifies an arithmetic locus, not a
physical mechanism. The absolute centered-KTA contrasts between
configurations are not deletion-stable
(\(\lvert z_{\mathrm{desc}}\rvert\le 0.87\)), and the uplift deletion
probe is likewise unresolved
(\(z_{\mathrm{desc}}\approx 1.1\)--\(1.2\)): the robust statement is the
fidelity ordering; the KTA ordering is a point-estimate pattern. We make
no claim that hardware improves supervised performance.

\hypertarget{finite-shot-reference-scales-rq3-and-synthesis}{%
\subsection{Finite-shot reference scales (RQ3) and
synthesis}\label{finite-shot-reference-scales-rq3-and-synthesis}}

The observed off-diagonal RMSE exceeds both finite-shot reference scales
in every configuration: \(0.0878\), \(0.0864\), and \(0.0427\) against
the conservative global scale
\(\sigma_{\mathrm{ref,global}}\approx 0.0221\) (about \(4.0\), \(3.9\),
and \(1.9\) global reference scales) and the matrix-aware scale
\(\approx 0.0083\) (about \(10.6\), \(10.5\), and \(5.0\) matrix-aware
scales). A direct reference simulation supports the same conclusion:
sampling-only reconstruction of the statevector matrix at the executed
shot count yields RMSE \(\approx 0.0082\) on average (\(99\)th
percentile \(\approx 0.0094\)) and full-matrix CKA \(\approx 0.9994\),
far below and above, respectively, the observed values. Under the stated
sampling-only reference models, finite-shot sampling alone therefore
cannot explain the observed hardware--statevector discrepancy. The
margin over the reference scales is narrowest for the gate-twirled job,
whose observed discrepancy is the smallest of the three; this is the
regime in which the originally planned 4096-shot budget would have
mattered most (the full quadrature bookkeeping, the fixed-RMSE
projection, and the dimensionless transforms are reported in
Supplementary Methods A9 and B3--B4; the matrix-aware quantities for
\texttt{H2} are pooled-binomial references that do not capture
between-randomization variability). The synthesis across RQ1--RQ4 is
then direct (Figure 2): the configuration that best matched the intended
statevector geometry did not show the largest observed hardware label
alignment; the fidelity leg of this cross-over is deletion-stable for
the Spearman, MAE, RMSE, and full-matrix CKA contrasts (on the
diagonal-excluded variant, the \texttt{M2}--\texttt{M0} contrast sits
just below the convention at \(z_{\mathrm{desc}}=1.95\), while the
\texttt{M2}--\texttt{M1} contrast at \(2.48\) passes it), while the KTA
leg is not resolved at all; and the observed discrepancy on which both
legs act exceeds the stated sampling-only reference scales in every
configuration.

\begin{figure}[!htbp]
\centering
\includegraphics[width=\linewidth]{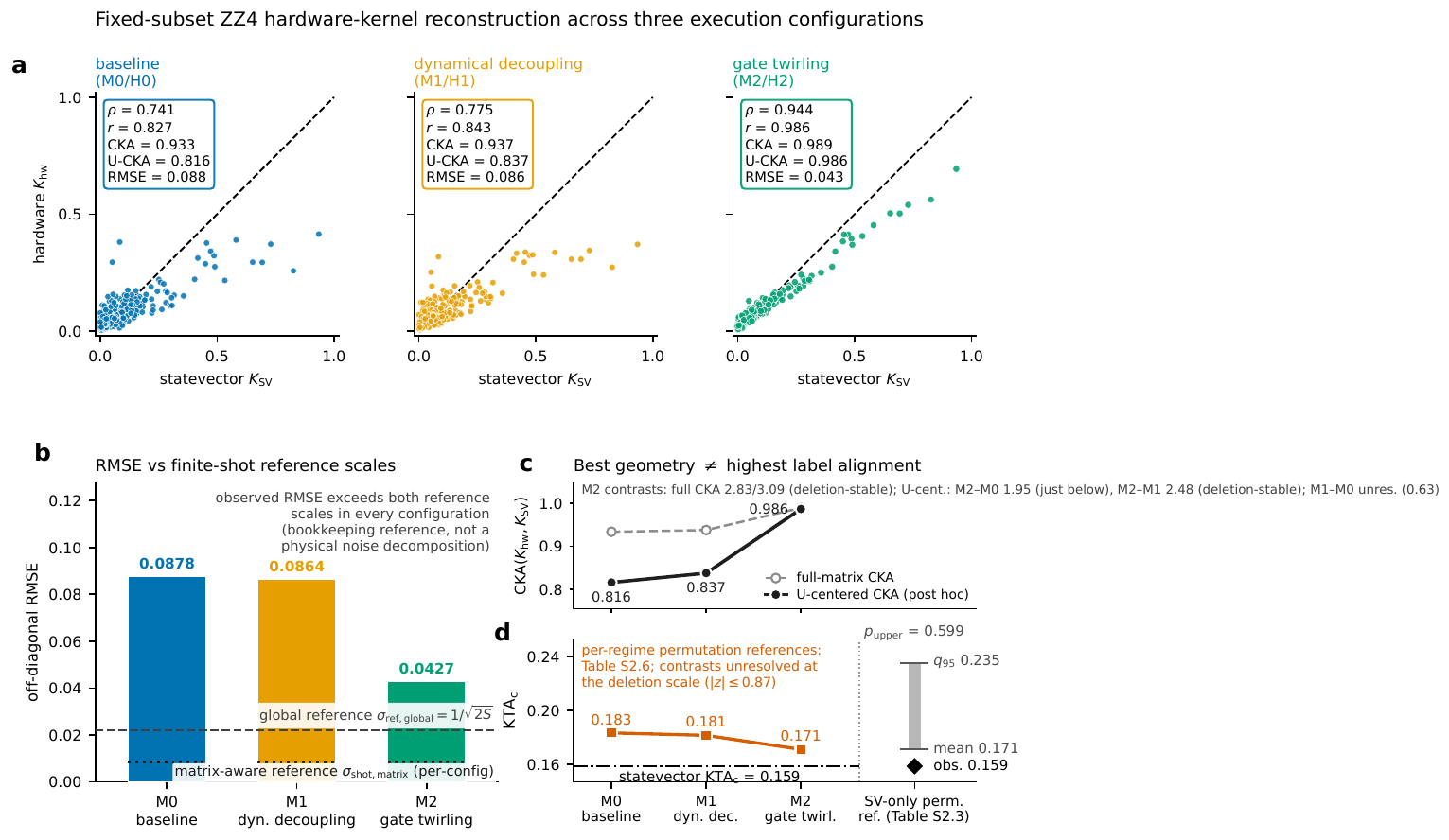}
\par\vspace{6pt}
\begin{minipage}{\linewidth}
\emph{Figure 2. Fixed-subset ZZ4 hardware-kernel reconstruction across
three execution configurations.} \emph{(a) Off-diagonal kernel entries
(276 unordered pairs) against the statevector reference; compression
toward a common background is strongest for baseline and weakest for
gate twirling. (b) Observed off-diagonal error against the global and
matrix-aware finite-shot reference scales; the observed RMSE exceeds
both in every configuration (quadrature bookkeeping, not a physical
noise-model decomposition). (c) Centered geometric fidelity (full-matrix
CKA) rises from \texttt{M0} to \texttt{M2}; the full-matrix \texttt{M2}
contrasts are deletion-stable (\(z_{\mathrm{desc}}=2.83\), \(3.09\)),
the diagonal-excluded \texttt{M2}--\texttt{M0} contrast sits just below
the convention (\(1.95\)) while \texttt{M2}--\texttt{M1} passes it
(\(2.48\)), and the \texttt{M1}--\texttt{M0} contrast is unresolved
(\(0.63\)). (d) Absolute hardware label alignment
(\(\operatorname{KTA}_{\mathrm{c}}\), point estimates; contrasts
unresolved, \(\lvert z\rvert\leq 0.87\)) falls as fidelity rises; the
grey interval is the statevector label-permutation reference (null mean
to \(q_{95}\)), and each hardware point sits at or below its own
random-label mean (Supplementary Table S2.6). Descriptive diagnostic on
the frozen \(N=24\) subset; single non-interleaved jobs on
\texttt{ibm\_fez} at 1024 shots.}
\end{minipage}
\end{figure}

\hypertarget{discussion}{%
\section{Discussion}\label{discussion}}

\hypertarget{what-the-case-study-establishes}{%
\subsection{What the case study
establishes}\label{what-the-case-study-establishes}}

The ZZ4 kernel was reconstructed on hardware in a limited, well-defined
sense: three complete, finite, already positive-semidefinite Gram
reconstructions, each retaining substantial rank-order, linear, and
centered agreement with the statevector reference, and each compressing
the off-diagonal spread as concentration theory anticipates
\citep{thanasilp2024exponential}. Among the three observed jobs, the
gate-twirled job was the most faithful on every reported axis, with
contrasts against both other jobs that are deletion-stable for Spearman,
MAE, RMSE, and full-matrix CKA (the Pearson \texttt{M2}--\texttt{M0} and
diagonal-excluded \texttt{M2}--\texttt{M0} contrasts fall just below the
convention); dynamical decoupling alone was not separated from baseline.
Because the configurations ran as single non-interleaved jobs on one
backend, the ordering is descriptive of these jobs: job-to-job variation
and calibration drift are confounded with configuration, and no causal
mitigation-efficacy estimate is made
\citep{wallman2016noise, viola1999dynamical, ji2024synergistic}.

\hypertarget{fidelity-and-task-relevance-are-distinct-axes}{%
\subsection{Fidelity and task relevance are distinct
axes}\label{fidelity-and-task-relevance-are-distinct-axes}}

The central lesson is the CKA/KTA tension: the reconstruction most
faithful to the designed kernel carried the lowest hardware label
alignment. The resolution is that the designed kernel was not aligned
with the task to begin with: its statevector alignment sits at or below
the random-label reference, as do the classical comparators. Greater
observed fidelity to the intended structure therefore did not coincide
with higher observed task alignment on this subset. The small uplift of
the noisier jobs is normalization-associated (stronger contraction of
the centered kernel norm than of the label-directed numerator); it lies
outside the finite-shot null while remaining below each kernel's own
permutation reference. It is a property of the distortion, not captured
signal. Implementation fidelity is necessary to realize a designed
quantum model, but it does not confer task relevance; on real data the
two axes need not coincide \citep{kubler2021inductive, huang2021power},
echoing the classical observation that similarity indices with different
invariances need not track task-grounded measures
\citep{raghu2017svcca, ding2021grounding, kornblith2019similarity}.
Hardware QML studies should report both axes; reporting either alone
invites overclaiming in one direction or the other.

\hypertarget{relation-to-concentration-spectral-alignment-and-kernel-design}{%
\subsection{Relation to concentration, spectral alignment, and kernel
design}\label{relation-to-concentration-spectral-alignment-and-kernel-design}}

Three strands of recent work bear on these results. First, kernel
concentration does not require hardware noise: the noiseless spectrum
and inductive bias of an embedding kernel are governed in part by
data-scaling (bandwidth-type) hyperparameters and by feature-map
architecture \citep{canatar2023bandwidth, gilfuster2024expressivity},
and in ZZ4 the scaling \(\alpha=2\) and the architecture (two
repetitions, linear entanglement) were fixed rather than tuned. The
hardware contraction measured here is therefore an \emph{additional}
distortion on top of whatever intrinsic concentration the fixed map
carries, and usefulness of the kernel is a separate question again
\citep{incudini2025toward}. Second, scalar centered KTA and spectral
task--model alignment are complementary, not interchangeable: with
\(K_c=V\Lambda V^{\top}\), the KTA numerator
\(y^{\top}K_c y=\sum_j \lambda_j (v_j^{\top}y)^2\) collapses the
eigenvalue-weighted label projection into one number, whereas the
spectral profile of Canatar, Bordelon, and Pehlevan
\citep{canatar2021spectral} resolves where target power lies across
eigenmodes and connects it to kernel-regression generalization. Neither
substitutes for statevector-referenced CKA, which measures
hardware-to-reference agreement rather than task alignment; a
quantitative spectral profile is beyond the frozen scope of this case
study, and kernel-regression generalization was not part of the design.
Third, an alternative to reproducing a fixed map more faithfully is to
adapt or discover the map itself \citep{incudini2022automatic}; within
this study the map is deliberately frozen, and any selection on the same
24 windows would be circular; a future design would select on training
data, freeze, and validate independently on hardware. Shot-budget-aware
acquisition \citep{xu2026aqka} is a further complementary direction for
the sampling side; it does not bear on the physical attribution of the
residual discrepancy, which this study deliberately leaves as a
reference-scale comparison.

\hypertarget{limitations}{%
\section{Limitations}\label{limitations}}

\textbf{Fixed subset and descriptive resolution.} All results rest on
one pre-authorized frozen subset of 24 windows from one duplicate-sensor
IAQ dataset and one binary target; the analysis supports
reconstruction-fidelity statements, not predictive ones. Uncertainty is
assessed only by leave-one-window-out deletion sensitivity;
\(z_{\mathrm{desc}}\) is not a significance test, non-resolution is not
equivalence, and median/maximum error, off-diagonal variance, and
effective rank carry no deletion probe at all. Nothing here licenses
extrapolation beyond this subset.

\textbf{Single backend, single non-interleaved jobs.} Each configuration
was executed once, non-interleaved, on \texttt{ibm\_fez} under one
pre-submission live backend-metadata snapshot (per-qubit calibration
tables were not persisted; Section 2.3), so the configuration comparison
is confounded with job-to-job variation and calibration drift. Because
the three jobs were submitted in a fixed order about eleven minutes
apart (Section 2.3), any drift in device conditions over that interval
is a competing explanation for the observed ordering that the recorded
data cannot exclude; the window-level jackknife quantifies deletion
sensitivity within a job, not variation between jobs. This is the most
consequential limit: the ordering is descriptive of these three jobs,
and no general mitigation-efficacy conclusion follows
\citep{ji2024synergistic}.

\textbf{Budget-safe shots and pooled twirling.} Execution used 1024
shots per circuit rather than the planned 4096; the reference-scale
conclusions hold under both budgets (Supplementary Methods B3--B4), at
the cost of precision where distortion is smallest. For \texttt{H2},
shots are pooled across 16 twirling randomizations; per-randomization
counts were not persisted, so between-randomization variability is
unestimated, and the binomial plug-in is a magnitude reference only. The
quadrature decompositions are deterministic bookkeeping identities, not
fitted noise models.

\textbf{Stopped revision-validation attempt.} A separate prospectively
specified technical validation attempt, not included in the Wave 1
results, stopped at its mandatory H2 acceptance gate because the tested
Runtime result representation (\texttt{ibm\_fez}) did not expose
separately addressable data for the 16 requested twirling
randomizations. No validation block was submitted and no primary
endpoint was computed; consequently, the attempt provides no replication
evidence and the Wave 1 H2 between-randomization variability remains
unestimated. This operational finding is specific to the tested software
stack and result-access path.

\textbf{One fixed map; no downstream evaluation; post hoc additions.} No
alternative encoding, bandwidth, or entanglement structure was explored,
and no classifier was trained: the endpoints are reconstruction fidelity
and distortion, not task performance \citep{singh2026benchmarking}. The
revision-added references (per-regime permutation nulls, resampling
null, attribution, U-centered CKA, deletion probes, comparators) are
deterministic or fixed-seed recomputations from the frozen artifacts,
added post hoc and marked as such; they supply reference context, not
configuration-level resolution.

\hypertarget{conclusion}{%
\section{Conclusion}\label{conclusion}}

This case study reported a fixed-protocol, statevector-referenced
measurement of how faithfully one frozen four-qubit ZZ4 quantum kernel
is reconstructed on a single IBM Quantum backend under three execution
configurations, each a single non-interleaved job. The intended
structure was reconstructed with characterizable,
configuration-dependent distortion: among the three observed jobs, the
gate-twirled job was the most faithful on every reported axis
(deletion-stable against the baseline for Spearman, MAE, RMSE, and
full-matrix CKA, and just below that convention for the Pearson and
diagonal-excluded variants), while dynamical decoupling alone was not
separated from baseline. The observed off-diagonal RMSE exceeds both
finite-shot reference scales in every configuration, so, under the
stated sampling-only reference models, finite-shot sampling alone does
not explain the discrepancy. The most faithful reconstruction did not
carry the highest label alignment, because the designed kernel was not
task-aligned to begin with; the small hardware alignment uplift is a
normalization property of the non-affine distortion, not captured
signal. Implementation fidelity and task relevance are distinct axes,
and hardware quantum machine-learning studies should report both. These
findings are descriptive and bounded to one kernel, one backend, three
jobs, and a frozen \(N=24\) subset; they support no claim of improved
prediction, hardware classifier superiority, or quantum advantage.
Converting them into causal and predictive statements is deferred to
future work under a new decision record, including interleaved and
replicated jobs, cross-device execution, per-randomization count
persistence, downstream classifier tests, and larger subsets. A future
validation study would require a separately amended acquisition
protocol, validated by a new technical sentinel, that preserves
per-randomization observations, followed by replicated time-bounded
blocks; such an acquisition design would not be assumed equivalent to
the historical server-side gate-twirling treatment.

\vspace{6pt}

\supplementary{Supplementary Note S1 (\emph{Artifact grounding and reproducibility
map}), Supplementary Note S2 (\emph{Extended statistical and diagnostic
tables}), Supplementary Note S3 (\emph{Extended figures}), and a
Supplementary Methods document (\emph{Extended methods, protocol, and
diagnostic detail}, relocated from the main text at revision) accompany
the online version of this article. Supplementary Note S1 provides the
consolidated section-to-artifact correspondence for the Methods and
Results, recorded against the round-0 section numbering (the mapping to
the revised structure is given in Supplementary Methods Table SM.0); it
documents the file paths, scripts, and checksum coverage that link each
frozen reported quantity to the persisted artifacts of the public Wave 1
reproducibility package, and its Section S1.5 reproduces the directional
expectations E1--E4 verbatim with their provenance status. Supplementary
Note S2 provides the full versions of the statistical and diagnostic
tables, together with the revision-added diagnostic tables
(Supplementary Tables S2.5--S2.9, each carrying its own provenance note
referencing package release \texttt{v1.3-revision-diagnostics}).
Supplementary Note S3 provides the extended figures: the finite-shot
reference-scale figure (Supplementary Figure S3.1) and the
spectral-flattening figure relocated from the main text (Supplementary
Figure S3.2, former Figure 3). The Supplementary Methods document
contains the relocated protocol, derivation, and extended diagnostic
material (Parts A1--A10 and B1--B6, with the location map in Table
SM.0).}

\authorcontributions{R.S. conceived and designed the study, developed the software, carried
out the experiments and the analysis, and wrote the manuscript. The author has read and agreed to the published version of the manuscript.}

\funding{This research received no external funding.}

\institutionalreview{Not applicable.}

\informedconsent{Not applicable.}

\dataavailability{The frozen ZZ4 hardware subset, the statevector reference kernel, and
the raw and reconstructed IBM Quantum result artifacts that support the
findings of this study are publicly available in the Wave 1
reproducibility repository at a fixed commit
(github.com/rsipakov/iaq-quantum-kernel-wave1-reproducibility, commit
\texttt{\seqsplit{6d14bca984486509b40850372f373c3499843dbc}}, release
tag \texttt{v1.2-wave1-manuscript}) and are archived at Zenodo (DOI:
10.5281/zenodo.21332398). This is an artifact-level reproducibility
package for the frozen ZZ4 hardware analysis. It does not reproduce the
full upstream indoor air-quality dataset construction. The audit and
analysis scripts used to reproduce the reported reconstruction,
distortion, uncertainty, and shot-noise diagnostics are openly available
in the same repository and archival snapshot
(github.com/rsipakov/iaq-quantum-kernel-wave1-reproducibility, commit
\texttt{\seqsplit{6d14bca984486509b40850372f373c3499843dbc}}, release
tag \texttt{v1.2-wave1-manuscript}; Zenodo DOI:
10.5281/zenodo.21332398), released under the MIT license. The
revision-added diagnostics of Section 2.5 are reproduced by the
fixed-seed scripts \texttt{scripts/09k\_revision\_diagnostics.py} and
\texttt{scripts/verify\_statevector\_regeneration.py}, included in
package release \texttt{v1.3-revision-diagnostics} of the same
repository (Zenodo DOI: 10.5281/zenodo.21438523), together with their
persisted output
(\texttt{hardware\_analysis/zz4\_wave1\_revision\_diagnostics.json}).
These scripts consume only artifacts byte-identical to the
\texttt{v1.2-wave1-manuscript} release, which remains the provenance
reference for all frozen quantities. The original numbered execution
scripts retained in the repository are archival records from the source
execution environment. They are not the supported reproduction path for
the flat public package.}

\acknowledgments{The author acknowledges the use of IBM Quantum services for this work.
The views expressed are those of the author and do not reflect the
official policy or position of IBM or the IBM Quantum team.

During the preparation of this manuscript, the author used Claude
(Anthropic) for the purposes of figure preparation, code-quality review,
readability editing, and proofreading; Perplexity for literature and
reference verification; and OpenAI Codex for an independent audit of the
reproducibility package. The author has reviewed and edited the output
of these tools and takes full responsibility for the content of this
publication.}

\conflictsofinterest{The author declares no competing interests.}

\begin{adjustwidth}{-\extralength}{0cm}
\reftitle{References}
\bibliography{quantum_kernel}

\begin{thebibliography}{999}

\bibitem[Havl{\'\i}{\v{c}}ek et~al.(2019)Havl{\'\i}{\v{c}}ek, C{\'o}rcoles,
  Temme, Harrow, Kandala, Chow, and Gambetta]{havlicek2019supervised}
Havl{\'\i}{\v{c}}ek, V.; C{\'o}rcoles, A.D.; Temme, K.; Harrow, A.W.; Kandala,
  A.; Chow, J.M.; Gambetta, J.M.
\newblock Supervised Learning with Quantum-Enhanced Feature Spaces.
\newblock {\em Nature} {\bf 2019}, {\em 567},~209--212.

\bibitem[Schuld and Killoran(2019)]{schuld2019quantum}
Schuld, M.; Killoran, N.
\newblock Quantum Machine Learning in Feature {Hilbert} Spaces.
\newblock {\em Physical Review Letters} {\bf 2019}, {\em 122},~040504.

\bibitem[Schuld(2021)]{schuld2021supervised}
Schuld, M.
\newblock Supervised Quantum Machine Learning Models Are Kernel Methods,  2021.
\newblock arXiv:2101.11020.

\bibitem[Thanasilp et~al.(2024)Thanasilp, Wang, Cerezo, and
  Holmes]{thanasilp2024exponential}
Thanasilp, S.; Wang, S.; Cerezo, M.; Holmes, Z.
\newblock Exponential Concentration in Quantum Kernel Methods.
\newblock {\em Nature Communications} {\bf 2024}, {\em 15},~5200.

\bibitem[Heyraud et~al.(2022)Heyraud, Li, Denis, Le~Boit{\'e}, and
  Ciuti]{heyraud2022noisy}
Heyraud, V.; Li, Z.; Denis, Z.; Le~Boit{\'e}, A.; Ciuti, C.
\newblock Noisy Quantum Kernel Machines.
\newblock {\em Physical Review A} {\bf 2022}, {\em 106},~052421.

\bibitem[Wang et~al.(2021)Wang, Du, Luo, and Tao]{wang2021towards}
Wang, X.; Du, Y.; Luo, Y.; Tao, D.
\newblock Towards Understanding the Power of Quantum Kernels in the {NISQ} Era.
\newblock {\em Quantum} {\bf 2021}, {\em 5},~531.

\bibitem[Morawska et~al.(2018)Morawska, Thai, Liu,
  et~al.]{morawska2018applications}
Morawska, L.; Thai, P.K.; Liu, X.;  et~al.
\newblock Applications of Low-Cost Sensing Technologies for Air Quality
  Monitoring and Exposure Assessment: How Far Have They Gone?
\newblock {\em Environment International} {\bf 2018}, {\em 116},~286--299.

\bibitem[Karagulian et~al.(2019)Karagulian, Barbiere, Kotsev,
  et~al.]{karagulian2019review}
Karagulian, F.; Barbiere, M.; Kotsev, A.;  et~al.
\newblock Review of the Performance of Low-Cost Sensors for Air Quality
  Monitoring.
\newblock {\em Atmosphere} {\bf 2019}, {\em 10},~506.

\bibitem[Peters et~al.(2021)Peters, Caldeira, Ho, Leichenauer, Mohseni, Neven,
  Spentzouris, Strain, and Perdue]{peters2021machine}
Peters, E.; Caldeira, J.; Ho, A.; Leichenauer, S.; Mohseni, M.; Neven, H.;
  Spentzouris, P.; Strain, D.; Perdue, G.N.
\newblock Machine Learning of High Dimensional Data on a Noisy Quantum
  Processor.
\newblock {\em npj Quantum Information} {\bf 2021}, {\em 7},~161.

\bibitem[Hubregtsen et~al.(2022)Hubregtsen, Wierichs, Gil-Fuster, Derks,
  Faehrmann, and Meyer]{hubregtsen2022training}
Hubregtsen, T.; Wierichs, D.; Gil-Fuster, E.; Derks, P.J.H.S.; Faehrmann, P.K.;
  Meyer, J.J.
\newblock Training Quantum Embedding Kernels on Near-Term Quantum Computers.
\newblock {\em Physical Review A} {\bf 2022}, {\em 106},~042431.

\bibitem[Glick et~al.(2024)Glick, Gujarati, C{\'o}rcoles, Kim, Kandala,
  Gambetta, and Temme]{glick2024covariant}
Glick, J.R.; Gujarati, T.P.; C{\'o}rcoles, A.D.; Kim, Y.; Kandala, A.;
  Gambetta, J.M.; Temme, K.
\newblock Covariant Quantum Kernels for Data with Group Structure.
\newblock {\em Nature Physics} {\bf 2024}, {\em 20},~479--483.

\bibitem[Huang et~al.(2021)Huang, Broughton, Mohseni, Babbush, Boixo, Neven,
  and McClean]{huang2021power}
Huang, H.Y.; Broughton, M.; Mohseni, M.; Babbush, R.; Boixo, S.; Neven, H.;
  McClean, J.R.
\newblock Power of Data in Quantum Machine Learning.
\newblock {\em Nature Communications} {\bf 2021}, {\em 12},~2631.

\bibitem[Schnabel and Roth(2025)]{schnabel2025quantum}
Schnabel, J.; Roth, M.
\newblock Quantum Kernel Methods Under Scrutiny: A Benchmarking Study.
\newblock {\em Quantum Machine Intelligence} {\bf 2025}, {\em 7},~58.

\bibitem[Gil-Fuster et~al.(2024)Gil-Fuster, Eisert, and
  Dunjko]{gilfuster2024expressivity}
Gil-Fuster, E.; Eisert, J.; Dunjko, V.
\newblock On the Expressivity of Embedding Quantum Kernels.
\newblock {\em Machine Learning: Science and Technology} {\bf 2024}, {\em
  5},~025003.

\bibitem[K{\"u}bler et~al.(2021)K{\"u}bler, Buchholz, and
  Sch{\"o}lkopf]{kubler2021inductive}
K{\"u}bler, J.M.; Buchholz, S.; Sch{\"o}lkopf, B.
\newblock The Inductive Bias of Quantum Kernels.
\newblock In Proceedings of the Advances in neural information processing
  systems (NeurIPS),  2021, Vol.~34, pp. 12661--12673.

\bibitem[Canatar et~al.(2023)Canatar, Peters, Pehlevan, Wild, and
  Shaydulin]{canatar2023bandwidth}
Canatar, A.; Peters, E.; Pehlevan, C.; Wild, S.M.; Shaydulin, R.
\newblock Bandwidth enables generalization in quantum kernel models.
\newblock {\em Transactions on Machine Learning Research} {\bf 2023}.
\newblock arXiv:2206.06686.

\bibitem[Incudini et~al.(2025)Incudini, Martini, and
  Di~Pierro]{incudini2025toward}
Incudini, M.; Martini, F.; Di~Pierro, A.
\newblock Toward useful quantum kernels.
\newblock {\em Advanced Quantum Technologies} {\bf 2025}, {\em 8},~2300298.
\newblock {\url{https://doi.org/10.1002/qute.202300298}}.

\bibitem[Ji and Polian(2024)]{ji2024synergistic}
Ji, Y.; Polian, I.
\newblock Synergistic Dynamical Decoupling and Circuit Design for Enhanced
  Algorithm Performance on Near-Term Quantum Devices.
\newblock {\em Entropy. An International and Interdisciplinary Journal of
  Entropy and Information Studies} {\bf 2024}, {\em 26},~586.

\bibitem[Cai et~al.(2023)Cai, Babbush, Benjamin, Endo, Huggins, Li, McClean,
  and O'Brien]{cai2023quantum}
Cai, Z.; Babbush, R.; Benjamin, S.C.; Endo, S.; Huggins, W.J.; Li, Y.; McClean,
  J.R.; O'Brien, T.E.
\newblock Quantum Error Mitigation.
\newblock {\em Reviews of Modern Physics} {\bf 2023}, {\em 95},~045005.

\bibitem[Viola et~al.(1999)Viola, Knill, and Lloyd]{viola1999dynamical}
Viola, L.; Knill, E.; Lloyd, S.
\newblock Dynamical Decoupling of Open Quantum Systems.
\newblock {\em Physical Review Letters} {\bf 1999}, {\em 82},~2417--2421.

\bibitem[Ezzell et~al.(2023)Ezzell, Pokharel, Tewala, Quiroz, and
  Lidar]{ezzell2023dynamical}
Ezzell, N.; Pokharel, B.; Tewala, L.; Quiroz, G.; Lidar, D.A.
\newblock Dynamical Decoupling for Superconducting Qubits: A Performance
  Survey.
\newblock {\em Physical Review Applied} {\bf 2023}, {\em 20},~064027.

\bibitem[Wallman and Emerson(2016)]{wallman2016noise}
Wallman, J.J.; Emerson, J.
\newblock Noise Tailoring for Scalable Quantum Computation via Randomized
  Compiling.
\newblock {\em Physical Review A} {\bf 2016}, {\em 94},~052325.

\bibitem[Temme et~al.(2017)Temme, Bravyi, and Gambetta]{temme2017error}
Temme, K.; Bravyi, S.; Gambetta, J.M.
\newblock Error Mitigation for Short-Depth Quantum Circuits.
\newblock {\em Physical Review Letters} {\bf 2017}, {\em 119},~180509.

\bibitem[Hicks et~al.(2022)Hicks, Kobrin, Bauer, and Nachman]{hicks2022active}
Hicks, R.; Kobrin, B.; Bauer, C.W.; Nachman, B.
\newblock Active Readout-Error Mitigation.
\newblock {\em Physical Review A} {\bf 2022}, {\em 105},~012419.

\bibitem[Smith et~al.(2021)Smith, Khosla, Self, and Kim]{smith2021qubit}
Smith, A.W.R.; Khosla, K.E.; Self, C.N.; Kim, M.S.
\newblock Qubit Readout Error Mitigation with Bit-Flip Averaging.
\newblock {\em Science Advances} {\bf 2021}, {\em 7},~eabi8009.

\bibitem[Kakavand et~al.(2026)Kakavand, Strohmeyer, and
  Schlotter]{kakavand2026benchmarking}
Kakavand, S.; Strohmeyer, C.; Schlotter, M.
\newblock Benchmarking Quantum Kernel Support Vector Machines Against Classical
  Baselines on Tabular Data: A Rigorous Empirical Study with Hardware
  Validation,  2026.
\newblock arXiv:2604.18837.

\bibitem[Singh et~al.(2026)Singh, Jin, and Merz~Jr.]{singh2026benchmarking}
Singh, G.; Jin, H.; Merz~Jr., K.M.
\newblock Benchmarking {MedMNIST} Dataset on Real Quantum Hardware.
\newblock {\em Scientific Reports} {\bf 2026}, {\em 16},~9017.

\bibitem[Cerezo et~al.(2022)Cerezo, Verdon, Huang, Cincio, and
  Coles]{cerezo2022challenges}
Cerezo, M.; Verdon, G.; Huang, H.Y.; Cincio, {\L}.; Coles, P.J.
\newblock Challenges and Opportunities in Quantum Machine Learning.
\newblock {\em Nature Computational Science} {\bf 2022}, {\em 2},~567--576.

\bibitem[Kornblith et~al.(2019)Kornblith, Norouzi, Lee, and
  Hinton]{kornblith2019similarity}
Kornblith, S.; Norouzi, M.; Lee, H.; Hinton, G.
\newblock Similarity of Neural Network Representations Revisited.
\newblock In Proceedings of the Proceedings of the 36th international
  conference on machine learning (ICML),  2019, Vol.~97, {\em PMLR}, pp.
  3519--3529.

\bibitem[Cortes et~al.(2012)Cortes, Mohri, and
  Rostamizadeh]{cortes2012algorithms}
Cortes, C.; Mohri, M.; Rostamizadeh, A.
\newblock Algorithms for Learning Kernels Based on Centered Alignment.
\newblock {\em Journal of Machine Learning Research} {\bf 2012}, {\em
  13},~795--828.

\bibitem[Roy and Vetterli(2007)]{roy2007effective}
Roy, O.; Vetterli, M.
\newblock The Effective Rank: A Measure of Effective Dimensionality.
\newblock In Proceedings of the 2007 15th european signal processing conference
  (EUSIPCO),  2007, pp. 606--610.

\bibitem[Higham(1988)]{higham1988computing}
Higham, N.J.
\newblock Computing a Nearest Symmetric Positive Semidefinite Matrix.
\newblock {\em Linear Algebra and its Applications} {\bf 1988}, {\em
  103},~103--118.

\bibitem[Rza(2026)]{rza2026beyond}
Rza, S.
\newblock Beyond Accuracy: A Kernel-Level Comparative Analysis of Quantum and
  Classical Support Vector Machines,  2026.

\bibitem[Shastry et~al.(2022)Shastry, Jayakumar, Patel, and
  Bhattacharyya]{shastry2022shot}
Shastry, A.; Jayakumar, A.; Patel, A.D.; Bhattacharyya, C.
\newblock Shot-Frugal and Robust Quantum Kernel Classifiers,  2022.
\newblock arXiv:2210.06971 (v3, 31 Dec 2023).

\bibitem[Javadi-Abhari et~al.(2024)Javadi-Abhari, Treinish, Krsulich, Wood,
  Lishman, Gacon, Martiel, Nation, Bishop, Cross, Johnson, and
  Gambetta]{javadi-abhari2024qiskit}
Javadi-Abhari, A.; Treinish, M.; Krsulich, K.; Wood, C.J.; Lishman, J.; Gacon,
  J.; Martiel, S.; Nation, P.D.; Bishop, L.S.; Cross, A.W.;  et~al.
\newblock Quantum Computing with {Qiskit},  2024.
\newblock arXiv:2405.08810.

\bibitem[Song et~al.(2012)Song, Smola, Gretton, Bedo, and
  Borgwardt]{song2012feature}
Song, L.; Smola, A.; Gretton, A.; Bedo, J.; Borgwardt, K.
\newblock Feature Selection via Dependence Maximization.
\newblock {\em Journal of Machine Learning Research} {\bf 2012}, {\em
  13},~1393--1434.

\bibitem[Efron and Stein(1981)]{efron1981jackknife}
Efron, B.; Stein, C.
\newblock The Jackknife Estimate of Variance.
\newblock {\em The Annals of Statistics} {\bf 1981}, {\em 9},~586--596.

\bibitem[Raghu et~al.(2017)Raghu, Gilmer, Yosinski, and
  Sohl-Dickstein]{raghu2017svcca}
Raghu, M.; Gilmer, J.; Yosinski, J.; Sohl-Dickstein, J.
\newblock SVCCA: {Singular} Vector Canonical Correlation Analysis for Deep
  Learning Dynamics and Interpretability.
\newblock In Proceedings of the Advances in neural information processing
  systems,  2017, Vol.~30.

\bibitem[Ding et~al.(2021)Ding, Denain, and Steinhardt]{ding2021grounding}
Ding, F.; Denain, J.S.; Steinhardt, J.
\newblock Grounding Representation Similarity with Statistical Testing.
\newblock In Proceedings of the Advances in neural information processing
  systems,  2021, Vol.~34.

\bibitem[Canatar et~al.(2021)Canatar, Bordelon, and
  Pehlevan]{canatar2021spectral}
Canatar, A.; Bordelon, B.; Pehlevan, C.
\newblock Spectral bias and task-model alignment explain generalization in
  kernel regression and infinitely wide neural networks.
\newblock {\em Nature Communications} {\bf 2021}, {\em 12},~2914.
\newblock {\url{https://doi.org/10.1038/s41467-021-23103-1}}.

\bibitem[Incudini et~al.(2022)Incudini, Lizzio~Bosco, Martini, Grossi, Serra,
  and Di~Pierro]{incudini2022automatic}
Incudini, M.; Lizzio~Bosco, D.; Martini, F.; Grossi, M.; Serra, G.; Di~Pierro,
  A.
\newblock Automatic and effective discovery of quantum kernels,  2022.
\newblock arXiv:2209.11144.

\bibitem[Xu et~al.(2026)Xu, Li, Zeng, Paisley, and Zhao]{xu2026aqka}
Xu, J.; Li, C.; Zeng, D.; Paisley, J.; Zhao, Q.
\newblock {AQKA}: Active quantum kernel acquisition under a shot budget,  2026.
\newblock arXiv:2605.14672.

\end{thebibliography}
%% \PublishersNote{} removed for the preprint build (publisher disclaimer)
\end{adjustwidth}
\end{document}